# Automodel solutions for superdiffusive transport by the Levy walks

A.B. Kukushkin[1,2,3], A.A. Kulichenko[1]

[1]National Research Center 'Kurchatov Institute', Moscow, 123182, Russian Federation
[2]National Research Nuclear University MEPhI, Moscow, 115409, Russian Federation
[3]Moscow Institute of Physics and Technology, Dolgoprudny, Moscow Region, 141700, Russian Federation

## Abstract

The method of approximate automodel solution for the Green's function of the time-dependent superdiffusive (nonlocal) transport equations (J. Phys. A: Math. Theor. 49 (2016) 255002) is extended to the case of a finite velocity of carriers. This corresponds to extension from the Lévy flights-based transport to the transport of the type, which belongs to the class of "Lévy walk + rests", to allow for the retardation effects in the Lévy flights. This problem covers the cases of the transport by the resonant photons in astrophysical gases and plasmas, heat transport by electromagnetic waves in plasmas, migration of predators, and other applications. We treat a model case of one-dimensional transport on a uniform background with a simple power-law step-length probability distribution function (PDF). A solution for arbitrary superdiffusive PDF is suggested, and the verification of solution for a particular power law PDF, which corresponds, e.g., to the Lorentzian wings of atomic spectral line shape for emission of photons, is carried out using the computation of the exact solution.

## 1. Introduction

A wide range of problems needs describing the transport in the medium for a finite velocity of carriers. The processes of nonlocal transport, which significantly differ from the conventional diffusion, are of special interest (see, e.g., the survey [1] and [2]). The energy transfer by photons in spectral lines of atoms and ions in plasmas and gases in astrophysical objects, nonlocal heat transport by electromagnetic waves in plasmas, migration of predators belong to such processes. These phenomena have superdiffusive character and have to be described by an integral equation in spatial coordinates, irreducible to a diffusion differential equation. The latter makes the numerical simulation of superdiffusive transport a formidable task.

The phenomenon of superdiffusion is closely related to the concept of Lévy flights [3-7]. The known example of such phenomenon is the radiative transfer in plasmas and gases in the Biberman-Holstein model [8-11]. This model considers resonance photon scattering by an atom or ion with complete redistribution over frequency in the act of absorption and re-emission. Here, rare distant flights of photons («jumps»), which correspond to emission/absorption in the «wings» of spectral line, dominate over contribution of frequent close displacements, which produce diffusive (Brownian) motion and correspond to emission/absorption in the core of spectral line. The distant flights caused by the long-tailed (e.g. power-law) wings of integral operator (i.e. of the step-length probability distribution function (PDF)) in the transport equation were shown [12] to be the Lévy flights. The dominant contribution of long-free-path photons to radiative transfer in spectral lines has been recognized in [13, 14]. The simple models based on this domination were developed for the quasi-steady-state transport, now known as the Escape Probability methods [15-17].

For the time-dependent superdiffusive transport by the Lévy flights, recently a wide class of the transport on a uniform background was shown [18-23] to possess an approximate automodel solution. The solutions for the Green's function were constructed using the scaling laws for the propagation front (i.e. time dependence of the relevant-to-superdiffusion average displacement of the carrier) and asymptotic solutions far beyond and far ahead the propagation front. The validity of the suggested automodel solutions was proved by their comparison with exact numerical solutions, in the one-dimensional (1D) case of the transport equation with a simple long-tailed PDF with various power-law exponents, and in the case of the Biberman-Holstein equation of the 3D resonance radiative transfer for various (Doppler, Lorentz, Voigt and Holtsmark) spectral line shapes.

The present work extends the method [18] to the case of a finite velocity of carriers. This corresponds to extension from the Lévy flights-based transport to the transport of the type, which belongs to the class of "Lévy walk + rests" (see Fig. 1 in [1]), to allow for the retardation effects in the Lévy flights. Similarly to [18], we treat a model case of one-dimensional transport on a uniform background with a simple power-law step-length PDF (section 2). A solution for arbitrary superdiffusive PDF is suggested (section 3), which uses the asymptotic solutions far beyond and far ahead the propagation front. The solution for particular power law PDF, which corresponds, e.g., to Lorentzian wings of atomic spectral line shapes for emission of photons and uses asymptotic solutions far beyond and far ahead the propagation front [24, 25], is presented in section 4. Its verification is performed in section 5, using the numerical simulation of the exact solution [24] of the transport equation. Modification of the solution and its verification are presented in sections 6.

## 2. Basic equation and general solution

The nonstationary equation for the Green's function $f(x,t)$ of the one-dimensional superdiffusive (nonlocal) transport of excitation in a homogeneous medium, with allowance for a finite velocity of the motion of carriers, has the form (derivation of this equation may be found in [24]):

$$\frac{\partial f(x,t)}{\partial t} = -\left(\frac{1}{\tau}+\sigma\right)f(x,t) + \frac{1}{\tau}\int_{-\infty}^{+\infty} dx' W(|x-x'|)f(x',t-\frac{|x-x'|}{c})\theta(t-\frac{|x-x'|}{c}) + \delta(x)\delta(t), \qquad (1)$$

where $W(\rho)$ is the step-length PDF, which describes the probability density for the process of carrier's start ("emission" of the carrier by the medium) and subsequent stop ("absorption" of the carrier by the medium) after passing the distance $\rho$,

$$\int_{-\infty}^{+\infty} W(|x-x'|)dx' = 1, \qquad (2)$$

$\tau$ is the average "waiting time", i.e. the time between the moments of stopping and starting the carrier (average lifetime of the medium's excitation), $c$ is the (constant) velocity of carriers, $\sigma$ is the average inverse lifetime of the carrier with respect to carrier's annihilation (deexcitation of medium); $\theta(x)$ is the Heaviside function; $\delta(x)$ is the Dirac delta-function.

Note that even for zero annihilation the volume-integrated excitation density is not conserved in time because the conservation law holds true only for the sum of volume-integrated values of medium's excitation density and carriers' density (if the transport problem is applied to the dynamics of the objects of the same type, for example, the search for food by animals, the total number of animals in motion and at rest will be the constant value).

For the dimensionless PDF

$$W(\rho) = 0.5\gamma/(1+\rho)^{\gamma+1}, \qquad 0 < \gamma < 2, \qquad (3)$$

general solution of Eq. (1) was obtained in [24]:

$$f(x,t,R_c) = \frac{1}{(2\pi)^2 i} \int_{-\infty}^{+\infty} dp\, e^{ipx} \lim_{\alpha \to +0} \int_{\alpha-i\infty}^{\alpha+i\infty} \frac{e^{st} ds}{s+1+\sigma\tau - \gamma \int_0^{+\infty} \frac{e^{-su/R_c} \cos(pu)}{(1+u)^{\gamma+1}} du}, \qquad (4)$$

where time $t$ is in the units of $\tau$, space coordinate $x$ is in the units of characteristic free path length $1/\kappa_0$ ($\kappa_0$ is the characteristic value of absorption coefficient), the retardation parameter $R_c = c\tau\kappa_0$ is the ratio of the average waiting time to the average time of flight. In what follows we consider the case $\sigma = 0$.

The asymptotic of the Green's function far ahead the propagation front was derived in [24]:

$$f(\rho \to tR_c - 0, t, R_c) = \left(t - \frac{\rho}{R_c}\right) W(\rho) \theta\left(t - \frac{\rho}{R_c}\right), \qquad \rho = |x|. \qquad (5)$$

In the case of infinite velocity of carriers ($R_c \to \infty$), it coincides with the respective asymptotics in [18] (see Eq. (6) therein).

## 3. Approximate automodel solution for arbitrary superdiffusive PDF

Following the principles of the method [18], we construct the following approximate automodel solution:

$$f_{auto}(x,t,R_c) = \left(t - \frac{\rho}{R_c} g\left(\frac{\rho_{fr}(t,R_c)}{\rho}\right)\right) W\left(\rho g\left(\frac{\rho_{fr}(t,R_c)}{\rho}\right)\right) \theta\left(t - \frac{\rho}{R_c} g\left(\frac{\rho_{fr}(t,R_c)}{\rho}\right)\right), \qquad \rho = |x|,$$
$$(6)$$

where the automodel function $g$ has the known asymptotic behavior,

$$g(s) = \begin{cases} 1, & s = s_{min} = \rho_{fr}(t,R_c)/(R_c t), \\ s, & s \gg s_{min}, \end{cases} \qquad (7)$$

where $\rho_{fr}(t,R_c)$ is the propagation front defined by the relation which equates the exact solutions in the alternative limits (cf. Eq. (25) in [23], instead of Eq. (5) in [18]), namely the asymptotics far ahead, Eq. (5), and the asymptotics far behind the propagation front, which is a plateau-like function, dependent on the time variable only:

$$\left(t - \frac{\rho_{fr}}{R_c}\right) W(\rho_{fr}) \theta\left(t - \frac{\rho_{fr}}{R_c}\right) = f(0,t,R_c). \qquad (8)$$

It is easy to prove that the solution (6)-(8) tends to the exact solution in the both alternative limits.

Similarly to [18], we introduce the function $Q$ needed for determination of the automodel function $g$ in the intermediate range of values of the automodel variable $s$:

$$\left(t - \frac{\rho}{R_c} Q(\rho, t, R_c)\right) W\left(\rho Q(\rho, t, R_c)\right) \theta\left(t - \frac{\rho}{R_c} Q(\rho, t, R_c)\right) = f(x, t, R_c). \tag{9}$$

To analyze the accuracy of the approximate automodel solution one has to show weak dependence of $Q_1$ and $Q_2$ functions on, respectively, space coordinate and time:

$$Q(\rho, t(\rho, s, R_c), R_c) \equiv Q_1(s, \rho, R_c) \simeq g(s, R_c), \tag{10}$$

$$Q(\rho(t, s, R_c), t, R_c) \equiv Q_2(s, t, R_c) \simeq g(s, R_c), \tag{11}$$

where the functions $t(\rho, s)$ and $\rho(t, s)$ are determined by the relation

$$s = \frac{\rho_{fr}(t, R_c)}{\rho}. \tag{12}$$

Note, that the definition (8) of the propagation front is a partial case of the definition of $\rho(t, s)$ with (8) and (12): $\rho_{fr}(t) = \rho(t, s = 1)$.

The verification of the automodel solution (6)-(8) should be done in the following way:
- calculation of the exact solution (4) in the range of space-time variable (not in the entire space of these variables) which will allow the determination of the automodel function in the range of values of $s \sim 1$, where the automodel solution is expected to be most sensitive to the interpolation between the limits (7) of high and low values of $s$,
- analysis of accuracy of self-similarity of the function $Q$, defined by Eq. (9), in the view of Eqs. (10) and/or (11),
- analysis of accuracy of the automodel solution (6)-(8) with respect to the exact solution (4).

## 4. Automodel solution for $\gamma=0.5$ power-law PDF

In what follows we consider the partial case $\gamma=0.5$ which corresponds, e.g., to the Lorentzian wings of atomic spectral line shape for emission of photons (see, e.g., the asymptotics of the Holstein function, Eq. (38) in [11]). The PDF (3) takes the form:

$$W(\rho) = \frac{1}{4(1+\rho)^{3/2}}. \tag{13}$$

Solving Eq. (8), we obtain explicit expression for the propagation front $\rho_{fr}(t, R_c)$:

$$\rho_{fr}(t,R_c) = \frac{1}{36R_c^2 f^2(0,t,R_c)} \left\{ \cos\left(\frac{\pi}{6} + \frac{1}{3}\arctg\left[\frac{\frac{1}{216R_c^2 f^2(0,t,R_c)} - 1 - tR_c}{\sqrt{(1+tR_c)\left(\frac{1}{108R_c^2 f^2(0,t,R_c)} - 1 - tR_c\right)}}\right]\right) - \frac{1}{2}\right\}^2 - 1.$$

(14)

The automodel function (9) may be expressed explicitly in terms of the automodel variable $s = \rho_{fr}(t,R_c)/\rho$:

$$Q_1(s,t,R_c) = Q(\rho_{fr}(t,R_c)/s, t, R_c) =$$

$$= \frac{s}{36\rho_{fr}(t,R_c)R_c^2 f_{exact}^2(\rho_{fr}(t,R_c)/s,t,R_c)} \left\{ \cos\left(\frac{\pi}{6} + \frac{1}{3}\arctg\left[\frac{\frac{1}{216R_c^2 f_{exact}^2(\rho_{fr}(t,R_c)/s,t,R_c)} - 1 - tR_c}{\sqrt{(1+tR_c)\left(\frac{1}{108R_c^2 f_{exact}^2(\rho_{fr}(t,R_c)/s,t,R_c)} - 1 - tR_c\right)}}\right]\right) - \frac{1}{2}\right\}^2 - \frac{s}{\rho_{fr}(t,R_c)}.$$

(15)

The automodel solution has the form

$$f_{auto}(x,t,R_c) = \left(t - \frac{\rho}{R_c}Q(\rho,t,R_c)\right)\frac{1}{4[1+\rho Q(x,t,R_c)]^{3/2}}, \qquad t > \frac{\rho}{R_c}Q(\rho,t,R_c). \tag{16}$$

The asymptotics far behind the propagation front for large retardation parameter and even larger values of time ($t \gg R_c \to \infty$) was calculated in [25]:

$$f(x \to 0, t \gg R_c, R_c \to \infty) = \frac{1}{R_c^{1/2} t^{3/2}} \frac{2}{\pi^2}\left[\frac{\pi}{8} + \frac{1}{4}\arccth\left(2\sqrt{2+\sqrt{3}}\right)\right] = \frac{0.0930}{R_c^{1/2} t^{3/2}}. \tag{17}$$

Equation (17) gives the scaling, which coincides with that of Eq. (19) in [26], and specifies the numerical coefficient.

## 5. Verification of automodel solution for $\gamma=0.5$ power-law PDF

We start the verification of the automodel solution (6)-(8) with the calculation of the exact solution (4) to determine of the automodel function in the range of values of $s\sim 1$, where the automodel solution is expected to be most sensitive to the interpolation between the limits (7) of high and low values of $s$.

Below are the results of numerical calculation of exact solution (4) for time moments $t=100, 300, 1000$ and retardation parameter $R_c=1, 10$.

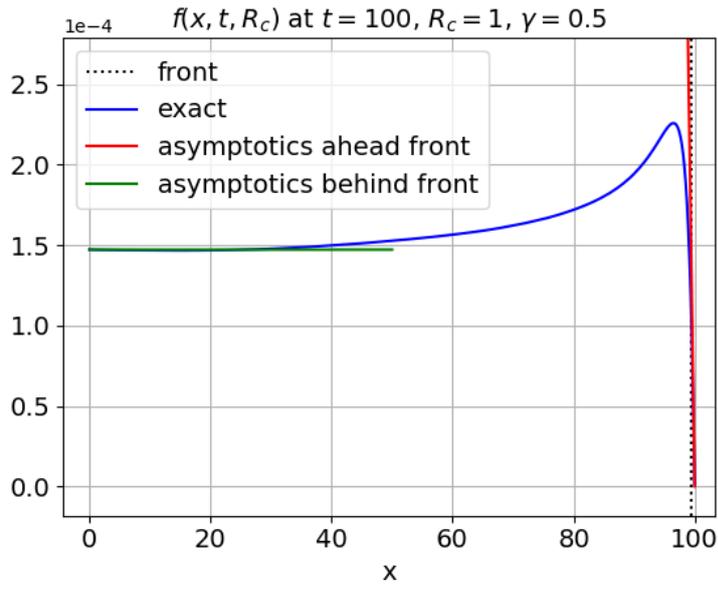

(a)

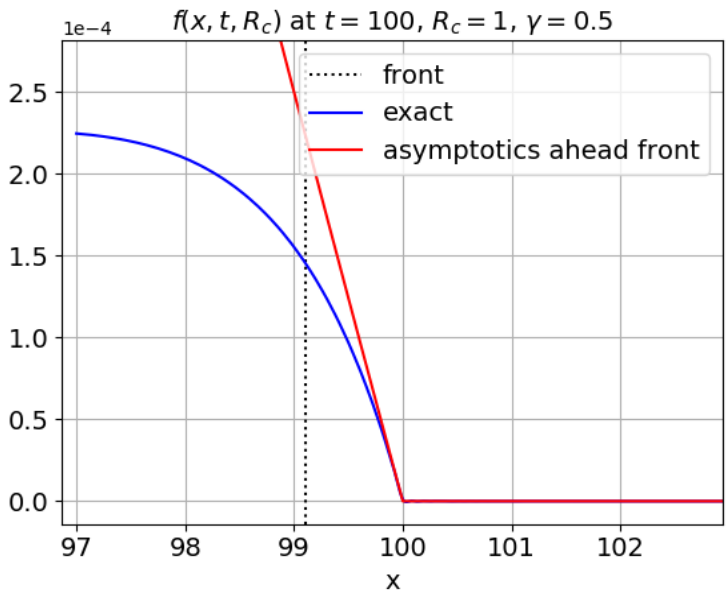

(b)

Figure 1. The result of the numerical calculation of the exact solution (4) for $t=100$ and $R_c=1$ in the entire space coordinate range (a) and near the ballistic cone (b). The asymptotics far ahead and far behind the propagation front, and the position of the propagation front are shown as well.

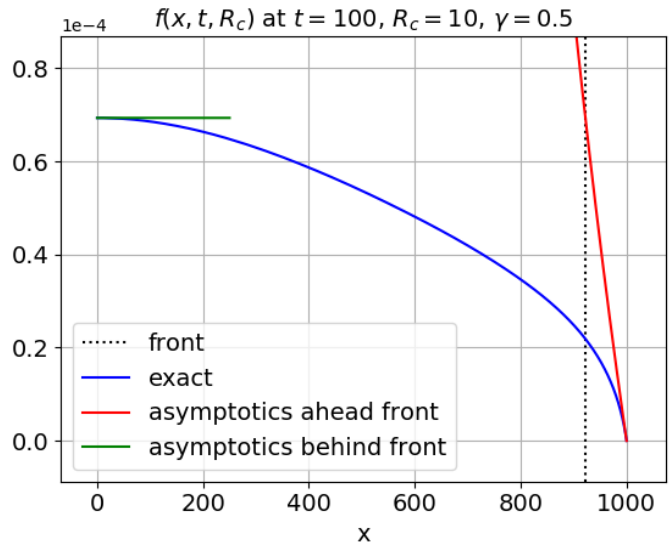

(a)

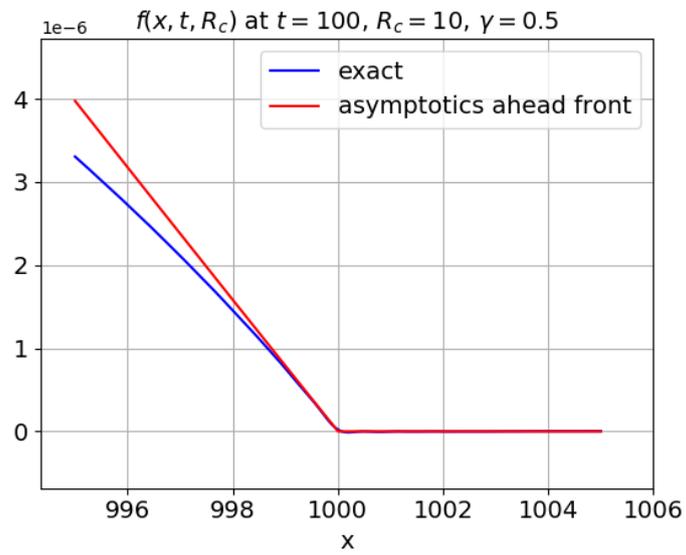

(b)

Figure 2. The same as in figure 1 but for $R_c$=10.

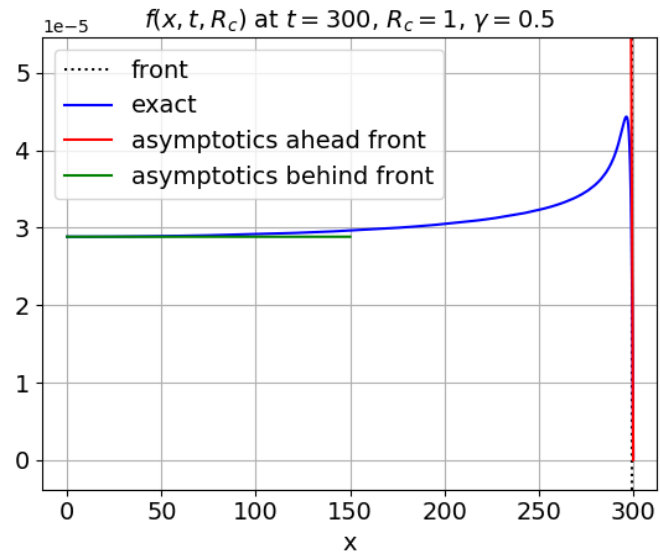

(a)

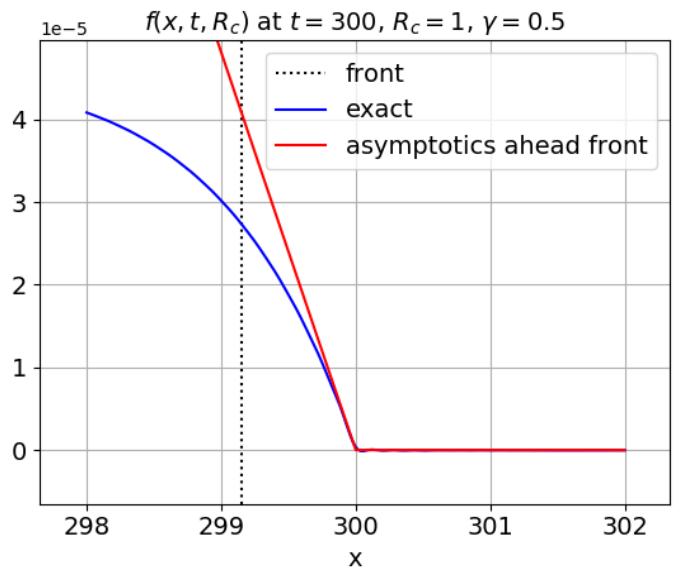

(b)

Figure 3. The same as in figure 1 but for $t=300$ and $R_c=1$.

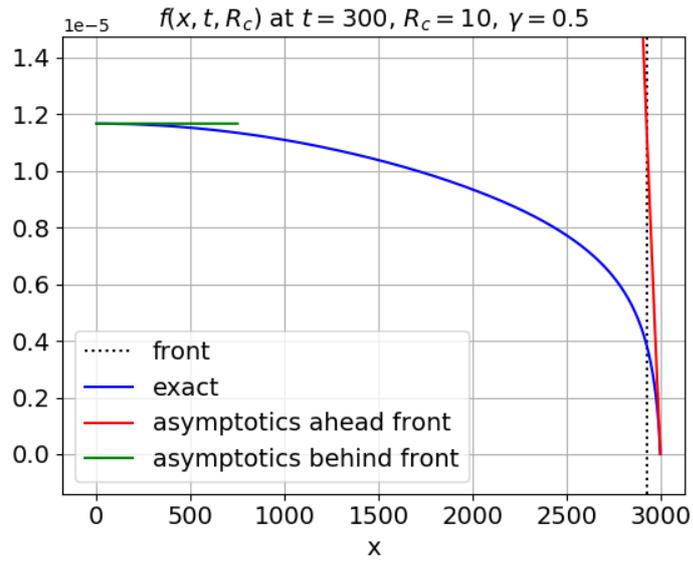

(a)

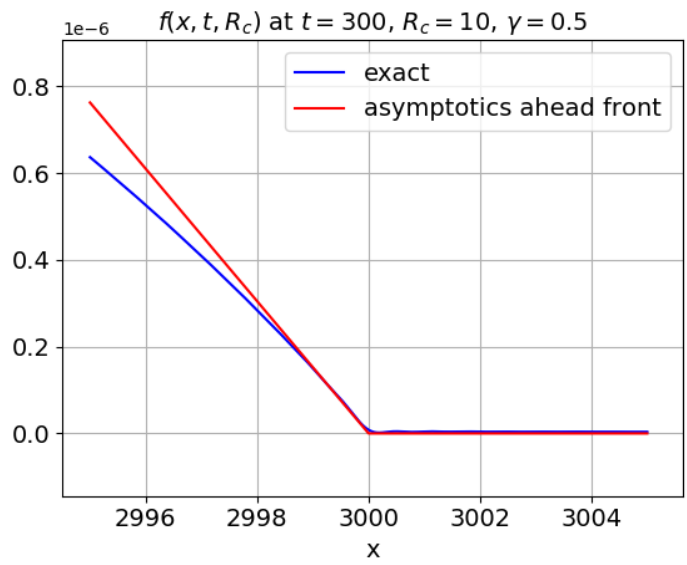

(b)

Figure 4. The same as in figure 1 but for $t=300$ and $R_c=10$.

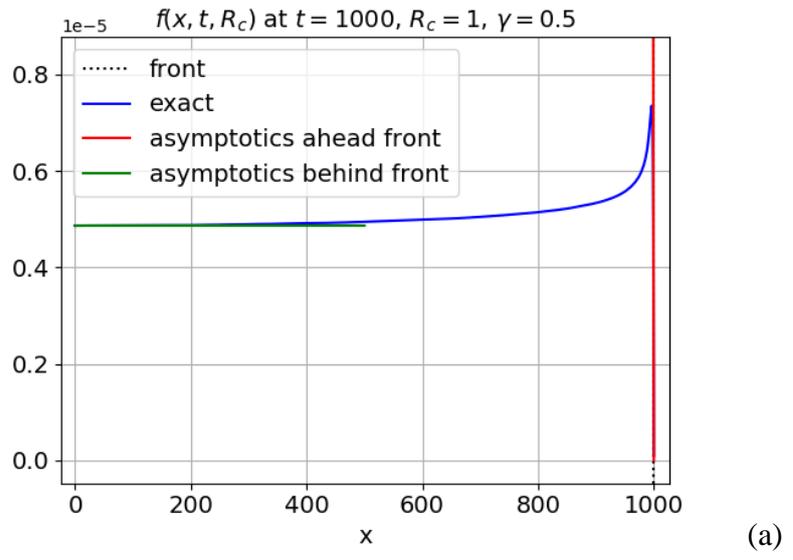

(a)

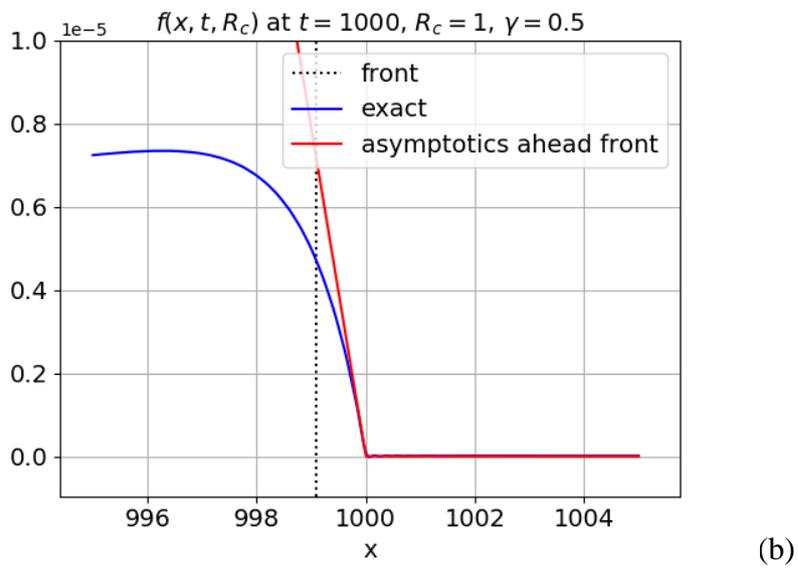

(b)

Figure 5. The same as in figure 1 but for $t$=1000 and $R_c$=1.

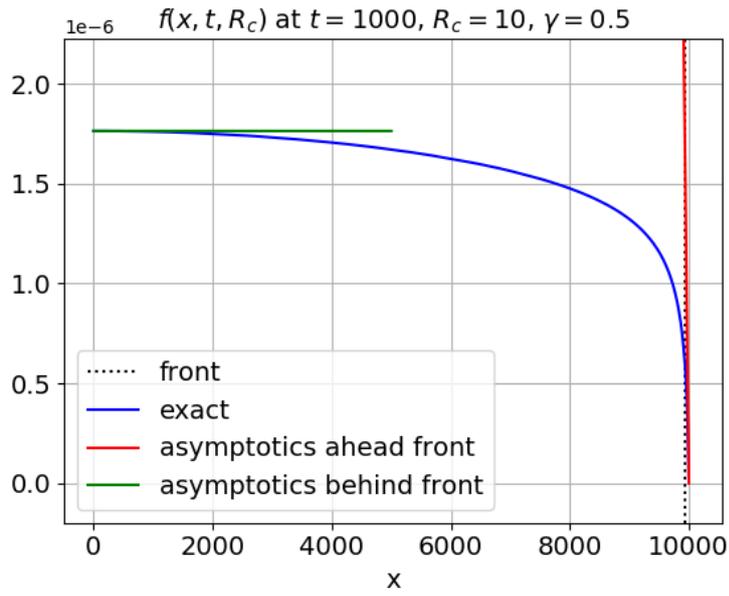

(b)

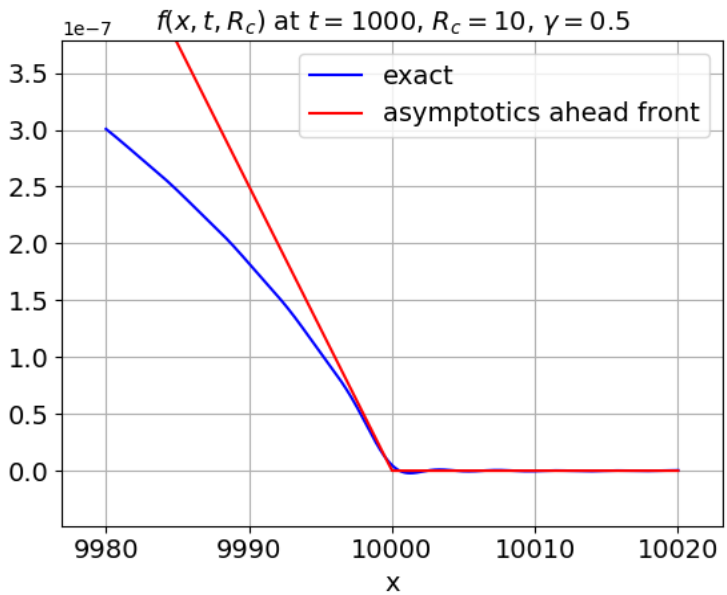

(b)

Figure 6. The same as in figure 1 but for $t=1000$ and $R_c=10$.

Using the results of calculating the exact solution, we can determine the function $Q$ (15) and analyze the accuracy of its self-similarity in the view of Eq. (11). The respective results are presented for $t=100$, 300, 1000, and $R_c=1$ (Figures 7, 8) and $R_c=10$ (Figures 9, 10).

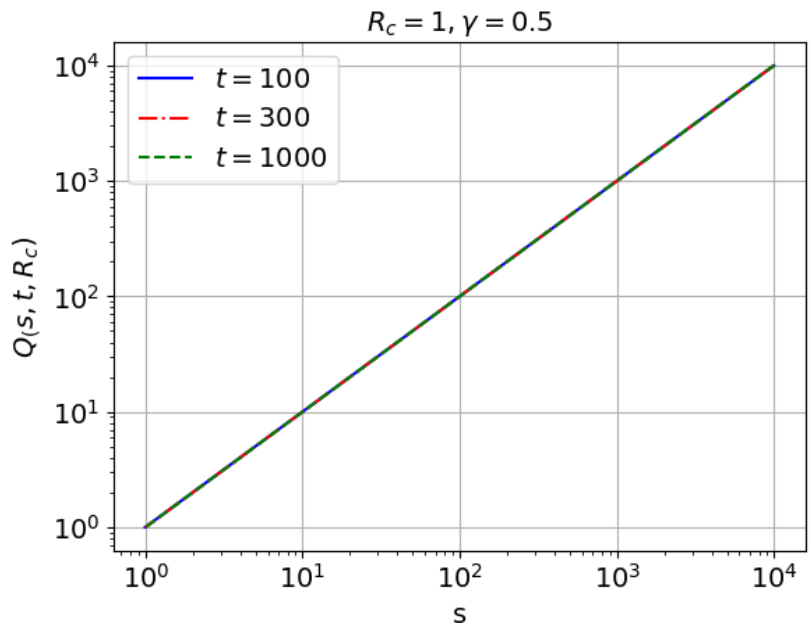

(a)

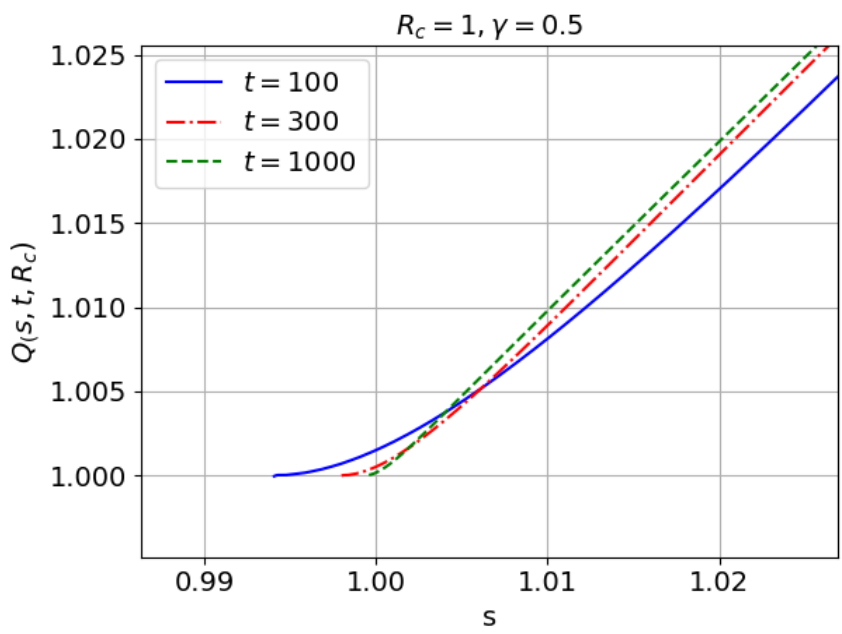

(b)

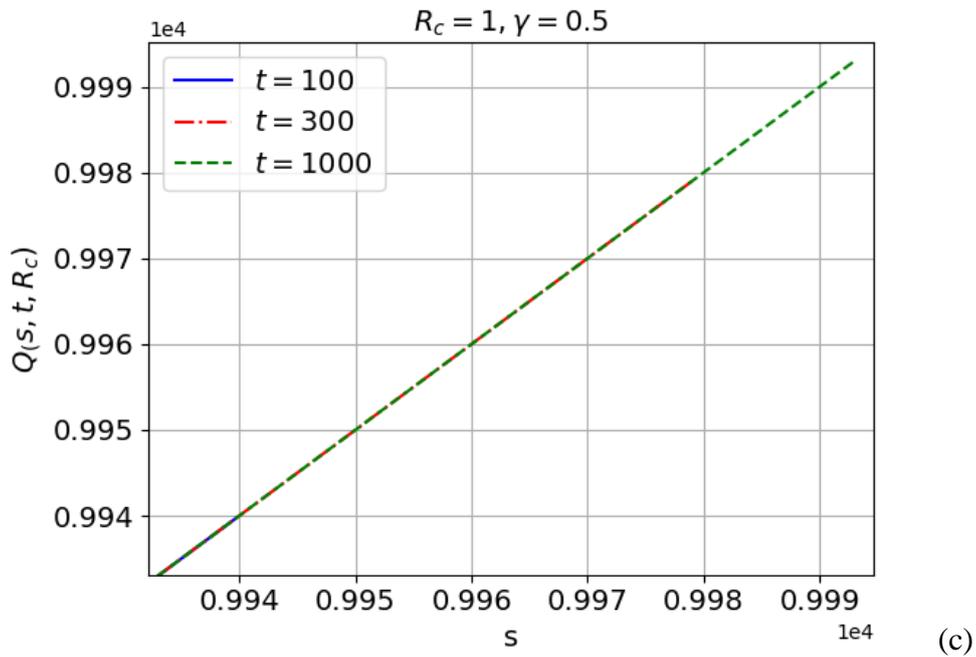

(c)

Figure 7. The function (15) of automodel variable s for $R_c$=1 and various time moments, $t$=100, 300, 1000, in various ranges of s.

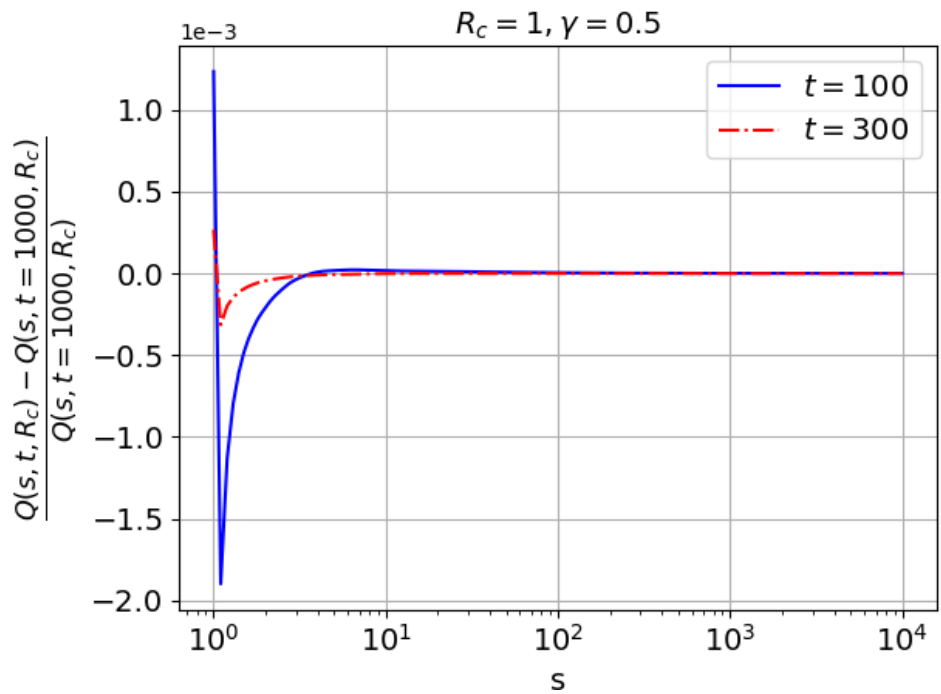

Figure 8. Characterization of self-similarity of the function (15) as a function of automodel variable s only, for $R_c$=1: relative deviation of (15) for $t$=100 and 300 from that for $t$=1000.

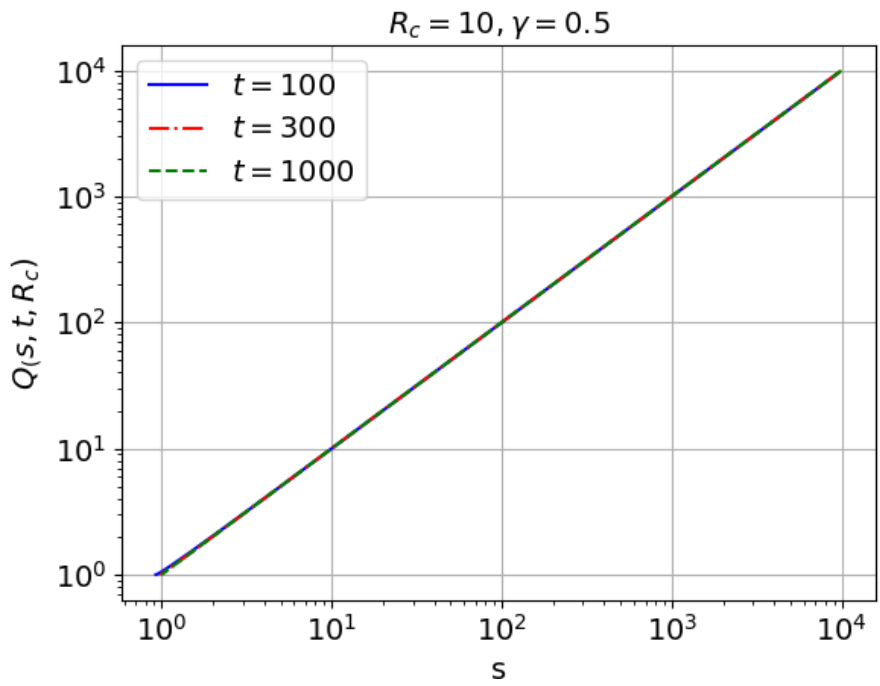

(a)

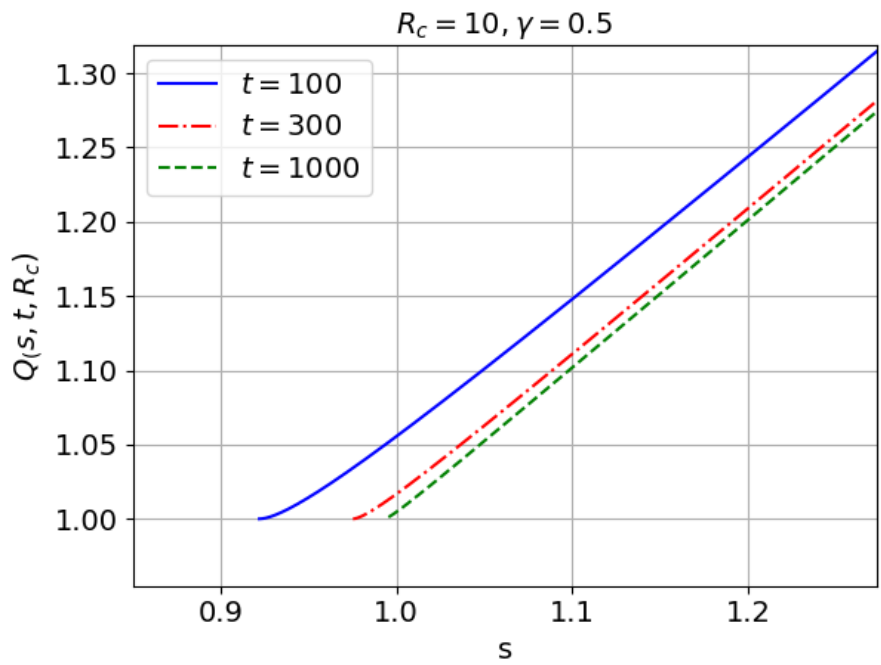

(b)

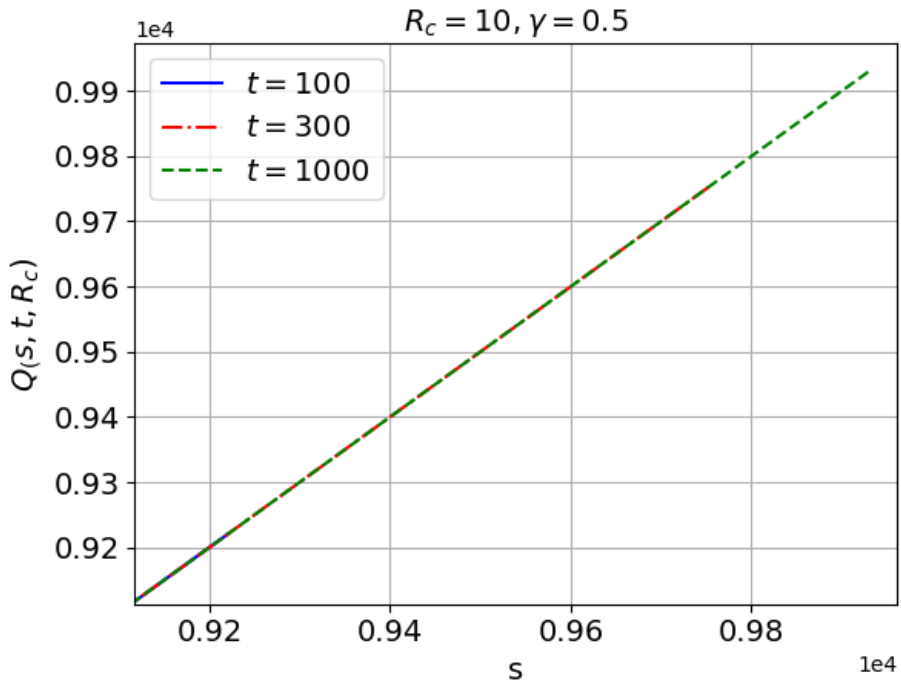

(c)

Figure 9. The same as in figure 7 but for $R_c=10$.

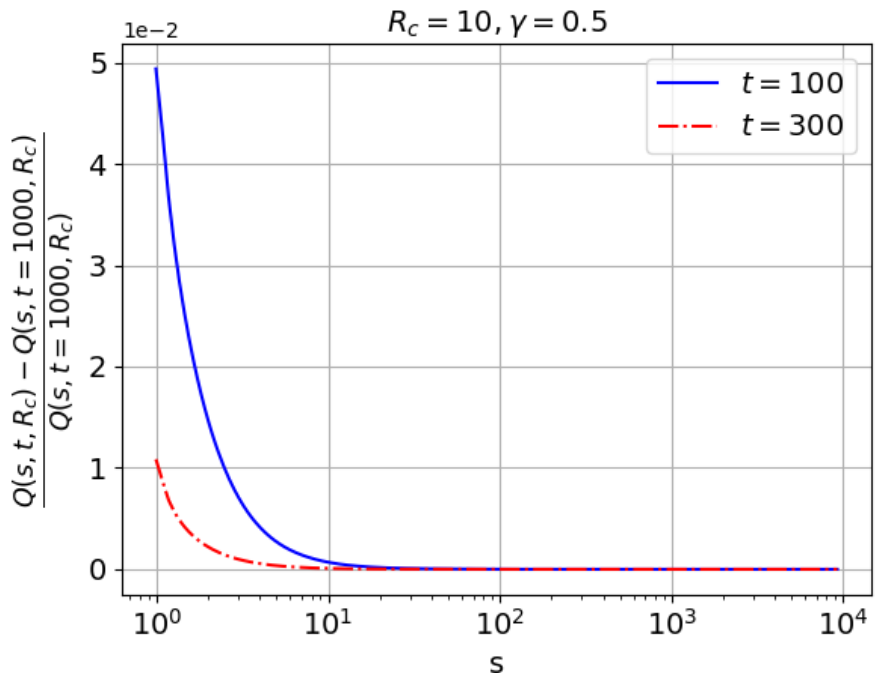

Figure 10. The same as in figure 8 but for $R_c=10$.

It is seen from Figures 7-10 that the accuracy of the self-similarity of the function (15), as a function of automodel variable *s* only (i.e. the accuracy of applicability of relation (11) which assumes independence of (15) on the time variable *t*), is high: relative deviation of (15) for $t=100$ and 300 from that for $t=1000$ does not exceed $2 \cdot 10^{-3}$ for $R_c=1$ and $5 \cdot 10^{-2}$ for $R_c=10$.

Now we are ready to make the next step: knowing the automodel function (15), we can construct the approximate automodel solution (6) and compare it with the calculated exact solution (4). The automodel solution for the Green's function is not applicable for simultaneously very small values of time and space coordinate: indeed, the automodel solution does not assume description of the evolution of the system immediately after the action of an instant point source. Therefore, for self-similarity analysis we will take the exact solution for $t=1000$ as a reference solution. This means that the automodel function (15), determined from comparison of exact and automodel solutions for $t=1000$, is used in the automodel solutions for other values of time. The respective comparison of automodel and exact solutions for $t=100$ and $300$ is presented in Figures 11(a) and 11(b), for $R_c=1$, and in Figures 12(a) and 12(b), for $R_c=10$.

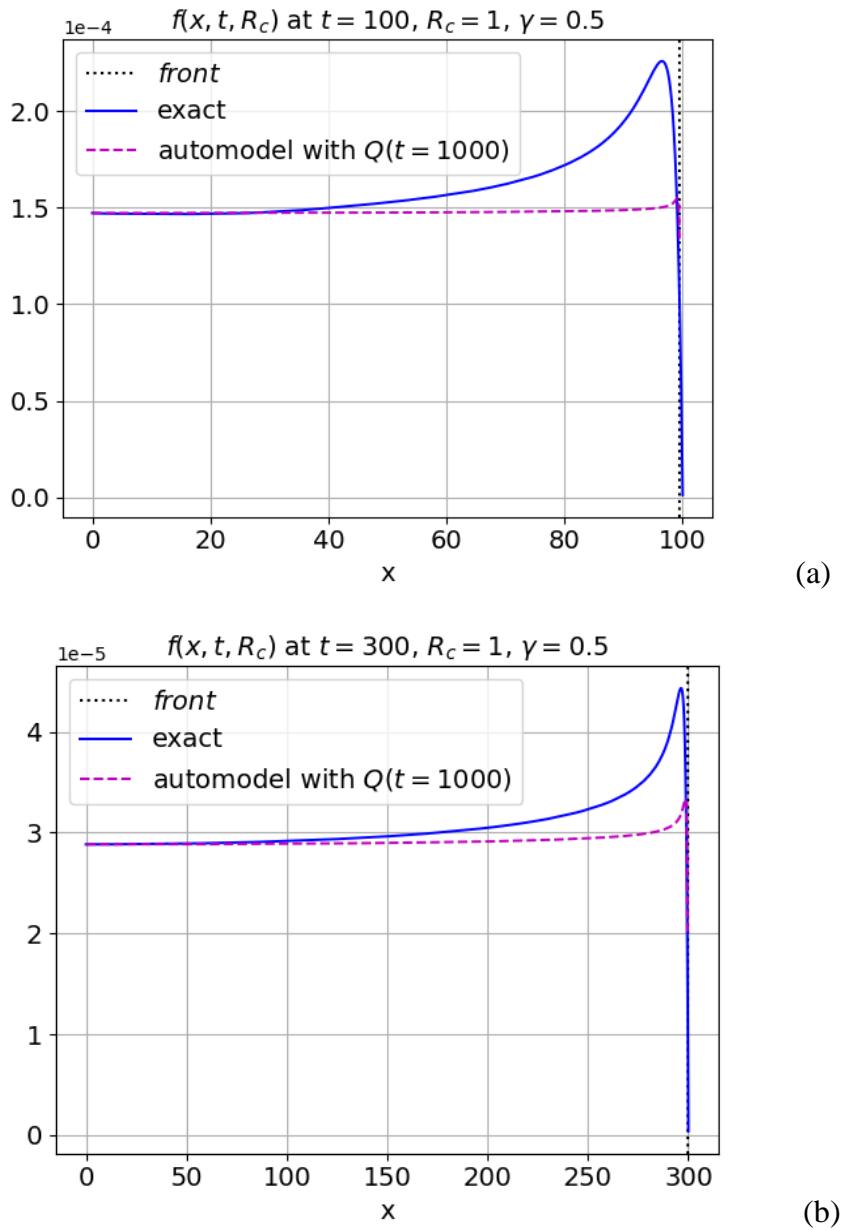

Figure 11. Comparison of automodel solution for $t=100$ (a) and $t=300$ (b) with automodel function $g(s)=Q(s, t=1000, R_c=1)$, recovered from comparison of exact and automodel solutions for $t=1000$, with the respective exact solution.

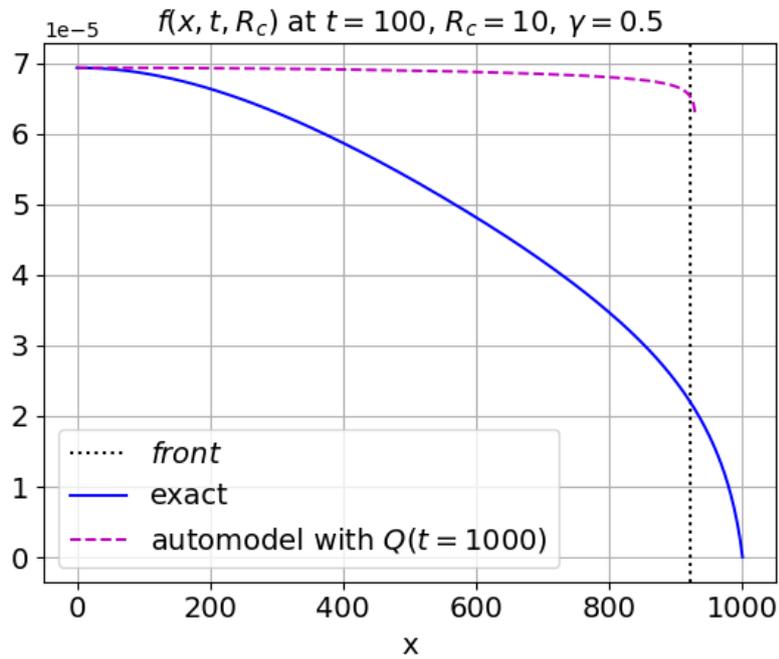

(a)

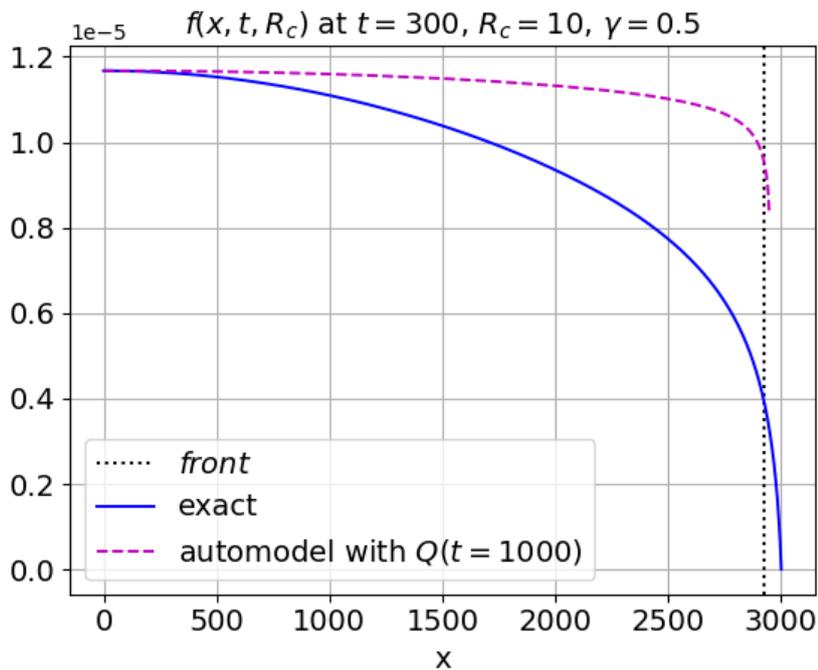

(b)

Figure 12. The same as in figure 11 but for $R_c = 10$.

It can be seen that, despite the small difference between the automodel functions (15) at various time moments, the deviation of the proposed approximate automodel solution (6) from the exact one is as high as a factor of few units in the region of the propagation front. Therefore, further modification of the approximate automodel solution (6) is necessary to improve the interpolation between the known asymptotics of the exact solution.

## 6. Modification of automodel solution for $\gamma=0.5$ power-law PDF

To improve the interpolation between the known asymptotics of the exact solution, we introduce a free parameter $\alpha$ in the definition of the propagation front (8):

$$\left(t - \frac{\rho_{fr}}{R_c}\right) W(\rho_{fr}) = \alpha f_{exact}(0, t, R_c). \tag{18}$$

For $W(\rho)$ (13), Eq. (18) gives the modification of the explicit expression (14) for the dependence of the propagation front on the time variable, which, in turn, determined the maximum allowable value of the parameter $\alpha$,

$$\alpha_{max}(t, R_c) = \frac{1}{R_c f_{exact}(0, t, R_c)\sqrt{108(1+tR_c)}} \tag{19}$$

and the respective value of the propagation front,

$$\lim_{\alpha \to \alpha_{max}} \rho_{fr}(t, R_c, \alpha) = \rho_{fr}(t, R_c, \alpha_{max}) = \frac{3}{4} tR_c - \frac{1}{4}. \tag{20}$$

It appears that the minimum of the logarithmic derivative

$$\frac{d \ln f_{auto}}{d \ln Q_1}(s, t, R_c, \alpha) = -\left[\frac{tR_c s}{\rho_{fr}(t, R_c, \alpha) Q_1(s, t, R_c, \alpha)} - 1\right]^{-1} - \frac{3}{2} \cdot \left[\frac{s}{\rho_{fr}(t, R_c, \alpha) Q_1(s, t, R_c, \alpha)} + 1\right]^{-1}, \tag{21}$$

in the range $s \sim 1$ is reached at $\alpha = \alpha_{max}(t, R_c)$.

The modification of the propagation front definition leads to a modification of the approximate automodel solution:

$$f_{auto}(x, t, R_c, \alpha_{max}(t, R_c)) = \left(t - \frac{\rho}{R_c} g\left(\frac{\rho_{fr}(t, R_c, \alpha_{max}(t, R_c))}{\rho}\right)\right) \times$$
$$\times W\left(\rho g\left(\rho_{fr}(t, R_c, \alpha_{max}(t, R_c))/\rho\right)\right) \frac{1}{\left\{\alpha_{max}(t, R_c) + (1 - \alpha_{max}(t, R_c))\frac{\rho}{R_c t}\right\}} \times \tag{22}$$
$$\times \theta\left(t - \frac{\rho}{R_c} g\left(\frac{\rho_{fr}(t, R_c, \alpha_{max}(t, R_c))}{\rho}\right)\right).$$

Where the asymptotics of $g(s)$ are not changed and defined by Eq. (7). The automodel function (15) is also modified to take the form

$$Q_1(s,t,R_c,\alpha) = Q(\rho_{fr}(t,R_c)/s,t,R_c,\alpha_{max}(t,R_c)) =$$

$$= \frac{s}{36\rho_{fr}(t,R_c,\alpha_{max}(t,R_c))R_c^2 f_{exact}^2(\rho_{fr}(t,R_c,\alpha_{max}(t,R_c))/s,t,R_c)\{\alpha_{max}(t,R_c)+(1-\alpha_{max}(t,R_c))s_{min}/s\}^2} \times$$

$$\times\left\{\cos\left(\frac{\pi}{6}+\frac{1}{3}\arctg\left[\frac{\frac{1}{216R_c^2 f_{exact}^2(\rho_{fr}(t,R_c,\alpha_{max}(t,R_c))/s,t,R_c)\{\alpha_{max}(t,R_c)+(1-\alpha_{max}(t,R_c))s_{min}/s\}^2}-1-tR_c}{\sqrt{(1+tR_c)\left(\frac{1}{108R_c^2 f_{exact}^2(\rho_{fr}(t,R_c,\alpha_{max}(t,R_c))/s,t,R_c)\{\alpha_{max}(t,R_c)+(1-\alpha_{max}(t,R_c))s_{min}/s\}^2}-1-tR_c\right)}}\right]\right)-\frac{1}{2}\right\}^2 -$$

$$-\frac{s}{\rho_{fr}(t,R_c,\alpha_{max}(t,R_c))}.$$

(23)

The dependence of the automodel function (23) on *s* for *t*=100, 300, 1000 and the characterization of the self-similarity of the function (23), as a function of automodel variable *s* only, are shown for $R_c$=10 in Figures 13 and 14, respectively, and for $R_c$=10 in Figures 15 and 16.

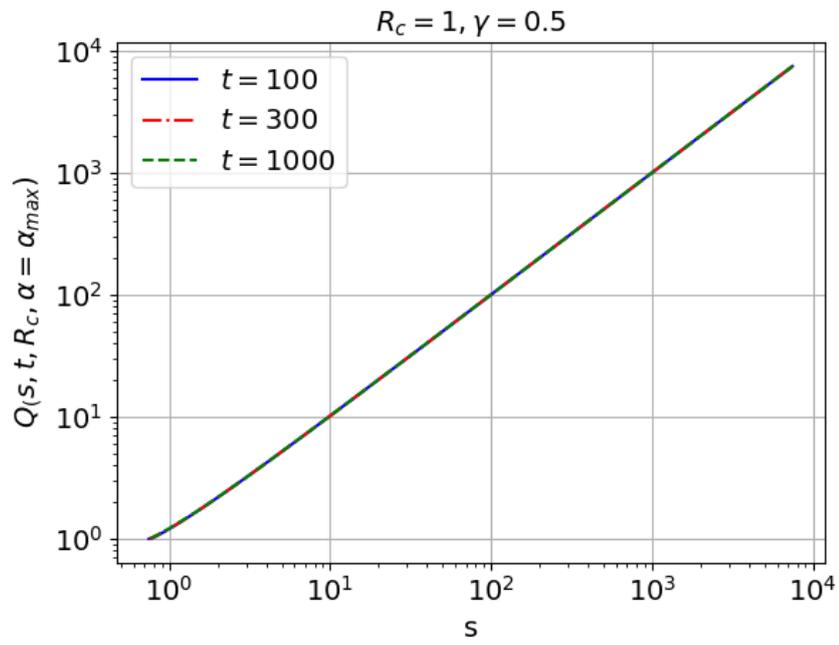

(a)

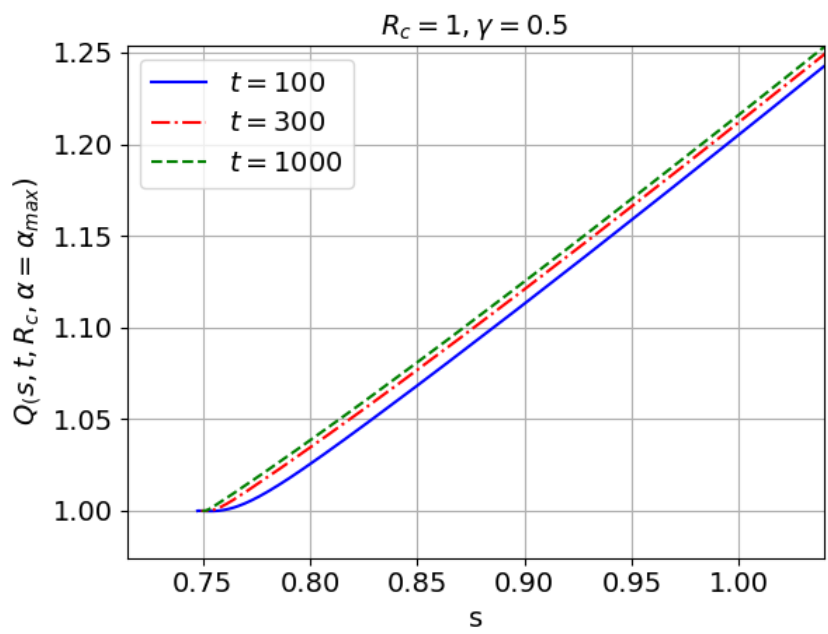

(b)

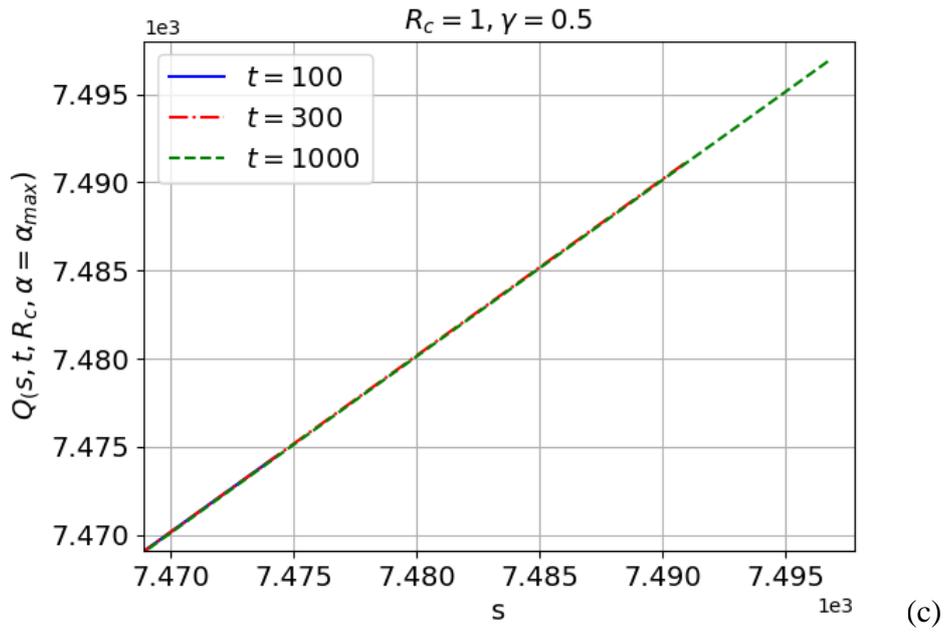

(c)

Figure 13. The function (23) of automodel variable s for $R_c$=1 and various time moments, $t$=100, 300, 1000, in various ranges of *s*.

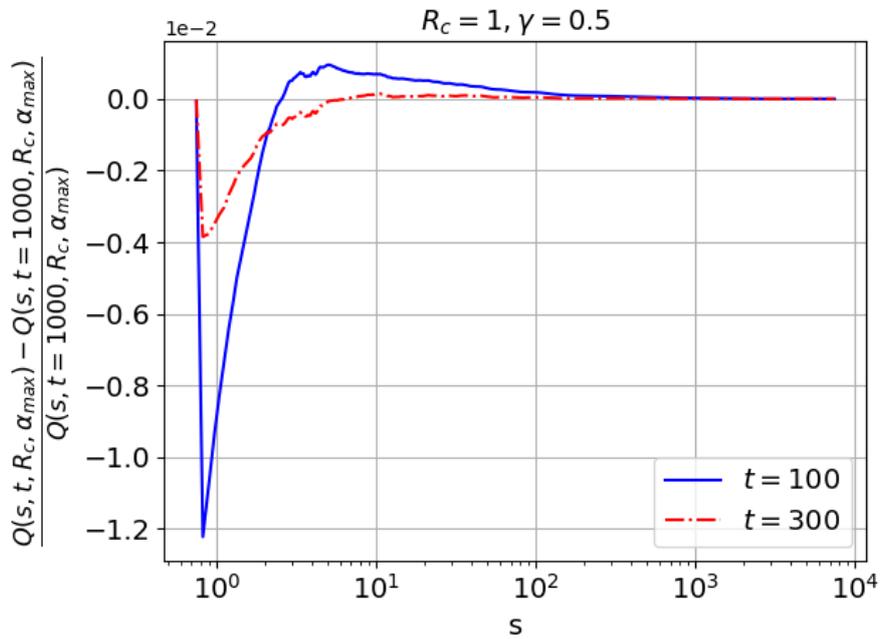

Figure 14. Characterization of self-similarity of the function (23) as a function of automodel variable *s* only, for $R_c$=1: relative deviation of (23) for $t$=100 and 300 from that for $t$=1000.

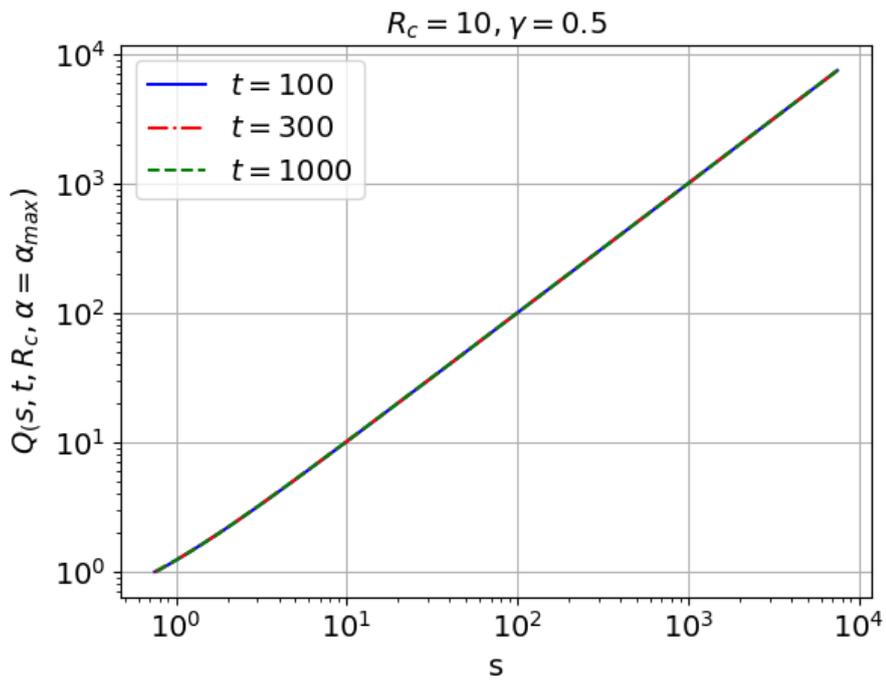

(a)

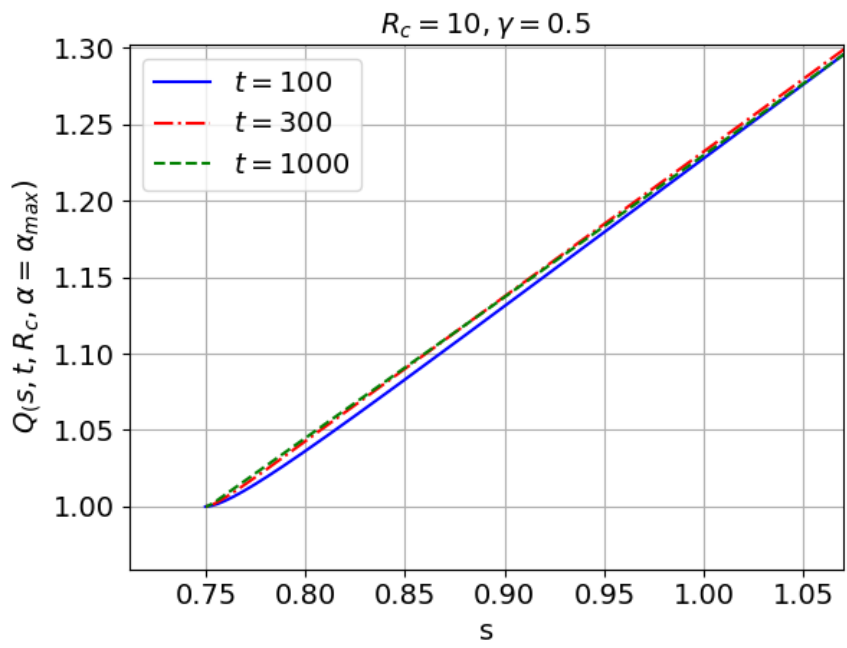

(b)

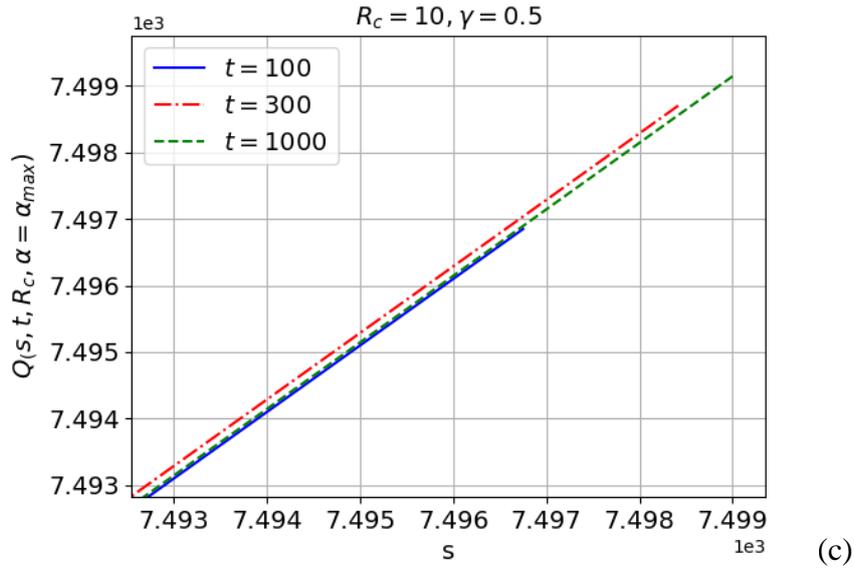

(c)

Figure 15. The same as in figure 13 but for $R_c=10$.

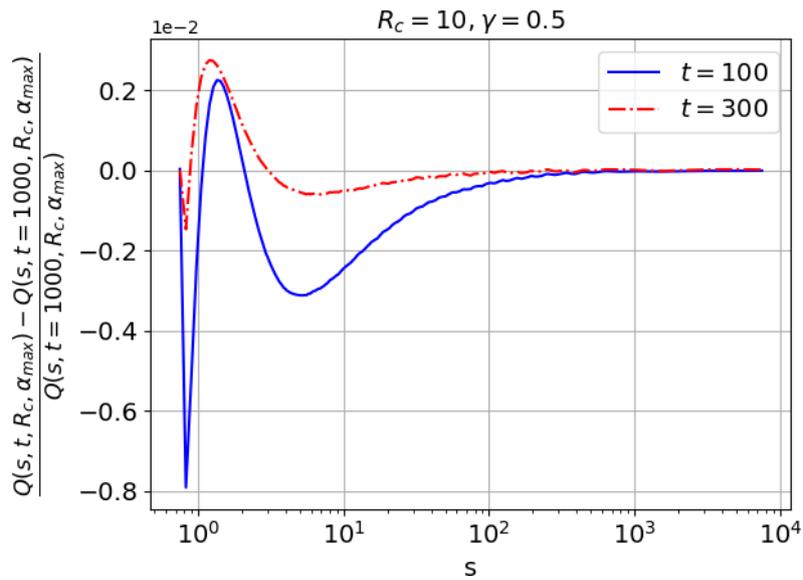

Figure 11. The same as in figure 14 but for $R_c=10$.

The final step of verifying the modified approximate automodel solution (22) is presented in Figures 17 and 18, quite similarly to Figures 11 and 12 for verification of the automodel solution (6). It appears that for $R_c=10$ we have substantial improvement of the accuracy of automodel solution, while for $R_c=1$ the accuracy is slightly worse. In general, the approximate automodel solution (22) seems to be better than that from Eq. (6).

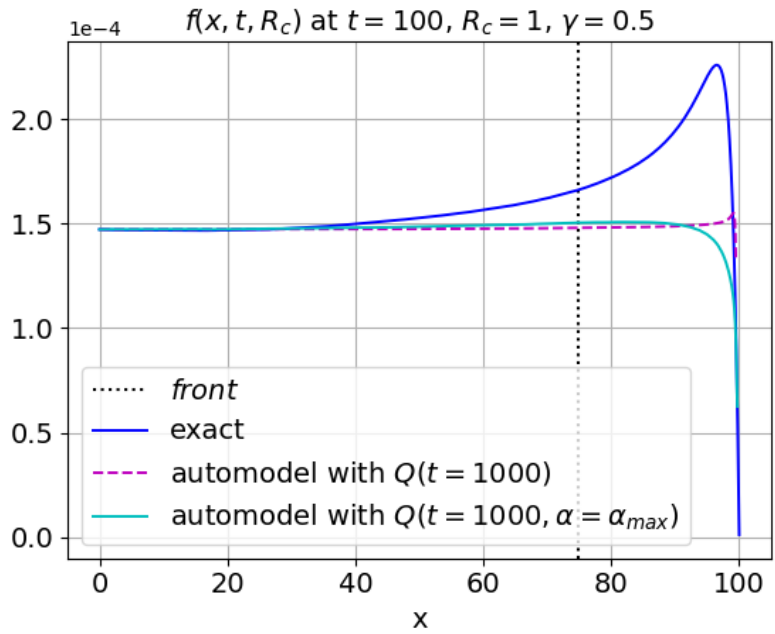

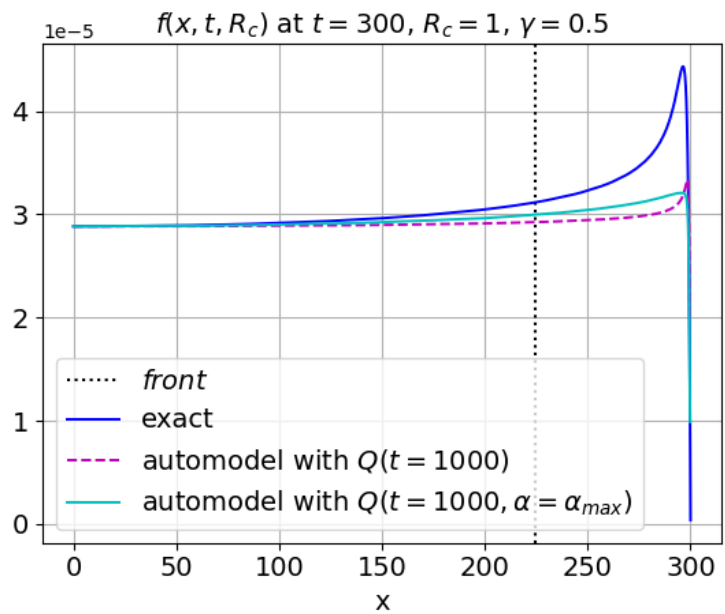

Figure 17. Comparison of exact solution (4) with two different approximate automodel solutions: automodel solution (6) with automodel function g(s)=Q(s, t=1000, $R_c$=1), where Q is given by Eq. (15) and recovered from comparison of exact (4) and automodel (6) solutions for t=1000; modified automodel solution (22) with automodel function g(s) = Q(s, t=1000, $R_c$=1,$\alpha_{max}$), where Q is given by Eq. (23) and recovered from comparison of exact (4) and automodel (22) solutions for t=1000 (the propagation front for automodel solution (22) is shown). Comparison is made for t=100 (a) and t=300 (b).

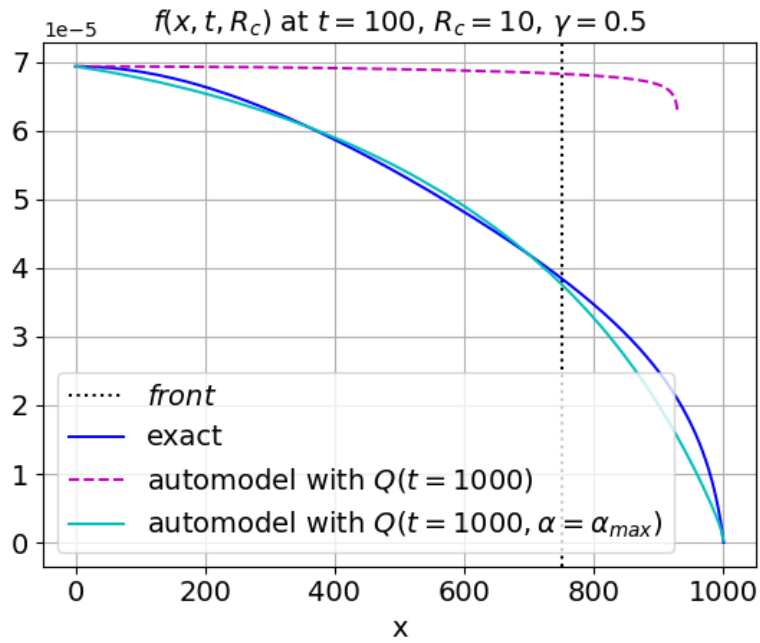

(a)

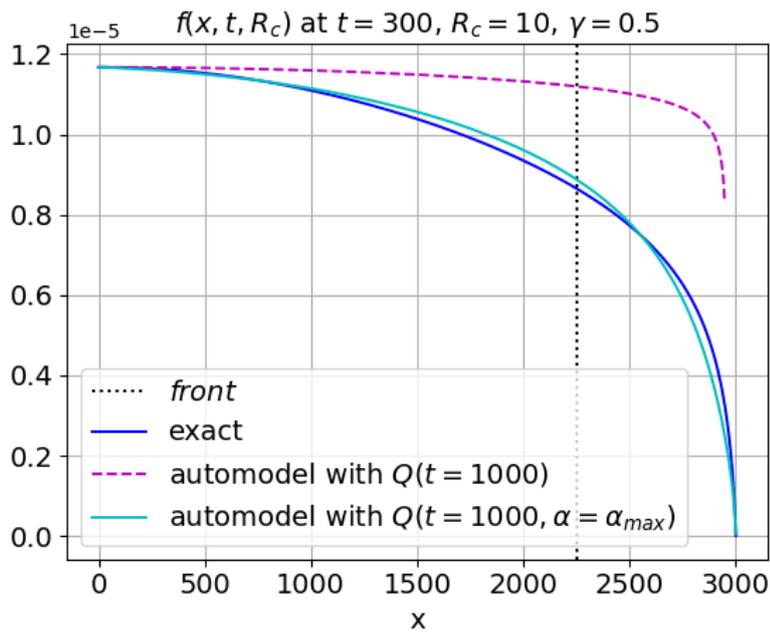

(b)

Figure 12. The same as in figure 17 but for $R_c$=10.

## 7. Conclusions

It is shown that the method [18] of approximate automodel solution for the Green's function of the time-dependent superdiffusive (nonlocal) transport on a uniform background may be extended to the case of a finite velocity of carriers. This corresponds to extension from the Lévy flights-based transport to the transport of the type, which belongs to the class of "Lévy walk + rests" [1] and is characterized by trajectories with a finite average waiting time and a finite velocity of carriers. A solution for arbitrary superdiffusive step-length probability distribution function (PDF) is suggested, and the verification of solution for a particular power law PDF, which corresponds, e.g., to the Lorentzian wings of atomic spectral line shape for emission of photons, is carried out using the computation of the exact solution. The success of identifying such solutions is based on the identification of the dominant role the long-free-path carriers in all three scaling laws used to construct the automodel solution, namely, the scaling laws for the propagation front (i.e. relevant-to-superdiffusion average displacement) and asymptotic solutions far beyond and far ahead the propagation front. The detailed analysis of automodel solutions for two values of the characteristic retardation parameter $R_c$ (the ratio of the average waiting time to the average time of flight) enabled us to evaluate the accuracy of defining the automodel function $g$ (i.e. accuracy of its self-similarity), which is not worse than few percent in a wide space-time region, and the resulting accuracy of the improved automodel solution, which is not worse than several tens of percent.

The results of the modification of the simplest automodel solution via modification of the propagation front in the framework of optimizing the parameters of interpolation between the known asymptotics of the exact solution show that there is a much freedom for such modifications to achieve the main goal of approximate automodel solutions for the Lévy flights-based transport and of various extensions of the principles [18], namely the construction of approximate solutions of superdiffusive transport problems with high enough accuracy with essential savings of computation time. Indeed, as shown in [22] and [23] for the case of Lévy flights transport, obtaining automodel (self-similar) solutions in the entire space of independent variables requires mass numerical simulations (distributed computing), however, their total volume is significantly reduced due to the self-similarity of the solution.

## Acknowledgements

The work is partly supported by the Russian Foundation for Basic Research (project RFBR 18-07-01269-a). This work has been carried out using computing resources of the federal collective usage center Complex for Simulation and Data Processing for Mega-science Facilities at NRC "Kurchatov Institute", http://ckp.nrcki.ru/.

The authors are grateful to K.V. Chukbar for helpful discussion of the paper [26].

## References

[1]  Zaburdaev V, Denisov S and Klafter J 2015 Lévy walks *Rev. Mod. Phys.* **87** 483
[2]  Shlesinger M F, Klafter J, and Wong J 1982 *J. Stat. Phys.* **27**, 499.
[3]  Mandelbrot B B 1982 *The Fractal Geometry of Nature* (New York: Freeman)
[4]  Shlesinger M, Zaslavsky G M and Frisch U (ed) 1995 *Lévy Flights and Related Topics in Physics* (New York: Springer)
[5]  Dubkov A A, Spagnolo B and Uchaikin V V 2008 Lévy flight superdiffusion: an introduction *Int. J. Bifurcation Chaos* **18** 2649
[6]  Klafter J and Sokolov I M 2005 Anomalous diffusion spreads its wings *Physics World* **18** 29
[7]  Eliazar I I and Shlesinger M F 2013 Fractional motions *Phys. Rep.* **527** 101–29
[8]  Biberman L M 1947 *Zh. Eksper. Teor. Fiz.* **17** 416; Biberman L M 1949 *Sov. Phys. JETP* **19** 584
[9]  Holstein T 1947 *Phys. Rev.* **72** 1212
[10] Biberman L M, Vorob'ev V S and Yakubov I T 1987 *Kinetics of Nonequilibrium Low Temperature Plasmas* (New York: Consultants Bureau)
[11] Abramov V A, Kogan V I and Lisitsa V S 1987 *Reviews of Plasma Physics* ed M A Leontovich and B B Kadomtsev vol 12 (New York: Consultants Bureau) p 151

[12] Pereira E, Martinho J and Berberan-Santos M 2004 Photon trajectories in incoherent atomic radiation trapping as Lévy flights *Phys. Rev. Lett.* **93** 120201
[13] Biberman L M 1948 *Dokl. Akad. Nauk SSSR* **49** 659
[14] Kogan V I 1968 *Proc. ICPIG'67: A Survey of Phenomena in Ionized Gases (Invited Papers)* (Russian: IAEA, Vienna) p 583
[15] Kalkofen W (ed) 1984 *Methods in Radiative Transfer* (Cambridge: Cambridge University Press)
[16] Rybicki G B *Ibid.* ch 1
[17] Napartovich A P 1971 *Teplofiz. Vys. Temp.* **9** 26
[18] Kukushkin A B and Sdvizhenskii P A 2016 Automodel solutions for Lévy flight-based transport on a uniform background *J. Phys. A: Math. Theor.* **49** 255002.
[19] Kukushkin A B and Sdvizhenskii P A 2014 Scaling Laws for Non-Stationary Biberman-Holstein Radiative Transfer *Proceedings of 41st EPS Conference on Plasma Physics, Berlin, Germany, 23 – 27 June 2014 ECA* **38F** P4.133 http://ocs.ciemat.es/EPS2014PAP/pdf/P4.133.pdf.
[20] Kukushkin A B, Sdvizhenskii P A, Voloshinov V V and Tarasov A S 2015 Scaling laws of Biberman-Holstein equation Green function and implications for superdiffusion transport algorithms *International Review of Atomic and Molecular Physics (IRAMP)* **6** 31–41 http://www.auburn.edu/cosam/departments/physics/iramp/6_1/kukushkin_et_al.pdf
[21] Kukushkin A B and Sdvizhenskii P A 2017 Accuracy analysis of automodel solutions for Lévy flight-based transport: from resonance radiative transfer to a simple general model. *J. Phys. Conf. Series* **941** 012050 http://iopscience.iop.org/article/10.1088/1742-6596/941/1/012050/pdf
[22] Kukushkin A B, Neverov V S, Sdvizhenskii P A and Voloshinov V V 2018 Numerical Analysis of Automodel Solutions for Superdiffusive Transport *International Journal of Open Information Technologies (INJOIT)* **6** 38–42 http://injoit.org/index.php/j1/article/view/535
[23] Kukushkin A B, Neverov V S, Sdvizhenskii P A and Voloshinov V V 2018 Automodel Solutions of Biberman-Holstein Equation for Stark Broadening of Spectral Lines *Atoms* **6(3)** 43 https://doi.org/10.3390/atoms6030043
[24] Kulichenko A A and Kukushkin A B 2017 *IRAMP* **8(1)** 5-14 http://www.auburn.edu/cosam/departments/physics/iramp/8_1/Kulichenko_Kukushkin.pdf.
[25] Kulichenko A A and Kukushkin A B 2018 *Proc. 45th EPS Conference on Plasma Phys., Prague, Chech Republic, 2018, ECA* **42A** P1.4013 http://ocs.ciemat.es/EPS2018PAP/pdf/P1.4013.pdf
[26] Zaburdaev V Yu, Chukbar K V 2002 *JETP* **94**(2) 252